\documentclass[english,prd,superscriptaddress,nofootinbib,preprintnumbers,showpacs,preprint]{revtex4}
\usepackage[latin1]{inputenc}
\usepackage{graphicx}
\usepackage{amsmath,amsthm,amssymb}
\usepackage{bm}

\usepackage{amsfonts}
\usepackage{dcolumn}

\usepackage{color}

\newcommand{\beqa}{\begin{eqnarray}}
\newcommand{\eeqa}{\end{eqnarray}}

\newcommand{\siml}{\lesssim}

\usepackage{babel}
\begin{document}

\title{Reconstructing the inflaton potential from the spectral index
}

\author{Takeshi Chiba}
\affiliation{Department of Physics, College of Humanities and Sciences, 
Nihon University, Tokyo 156-8550, Japan}

\begin{abstract}
Recent cosmological observations are in good agreement with 
the scalar spectral index $n_s$ with  $n_s-1\sim -2/N$, where $N$ is the number of e-foldings.  
Quadratic chaotic model, Starobinsky model and  Higgs inflation  or   
$\alpha$-attractors connecting them are typical examples  predicting such a relation. 
We consider the problem in the opposite: given $n_s$ as a function of $N$,
what is the inflaton potential $V(\phi)$.  We find that for $n_s-1=-2/N$, $V(\phi)$ is 
either $\tanh^2(\gamma\phi/2)$ ("T-model")  
or  $\phi^2$ (chaotic inflation) to the leading order in the slow-roll approximation.  
$\gamma$  is the ratio of $1/V$ at $N\rightarrow \infty$ to 
the slope of $1/V$ at a finite $N$ and is related to "$\alpha$" in the  
$\alpha$-attractors by $\gamma^2=2/3\alpha$.  
The tensor-to-scalar ratio $r$ is $r=8/N(\gamma^2 N +1) $.  
The implications for the reheating temperature are also discussed. 
We also derive formulas for $n_s-1=-p/N$.  We find that if the potential is 
bounded from above, only $p>1$ is allowed. 
Although $r$ depends on a parameter, the running of the spectral index is independent of it, 
which can be used as a consistency check of the assumed relation of $n_s(N)$. 
\end{abstract}

\date{\today}


\maketitle

\section{Introduction}

The latest Planck data \cite{Planck:2015} are in good agreement with 
the scalar spectral index $n_s$ with  $n_s-1\sim -2/N$, where $N$ is the number of e-foldings. 
Quadratic chaotic inflation model \cite{chaotic}, Starobinsky model \cite{alex} and Higgs inflation with a nonminimal coupling \cite{higgs}   
or $\alpha$-attractor  connecting them with one parameter "$\alpha$" \cite{kl,kl1,kl2} are typical examples  which predict such a relation. 
What else are there any inflation models predicting such a relation?  
In this note, we consider such an inverse problem :  
reconstruct $V(\phi)$ from a given $n_s(N)$.   
In Sec.\ref{sec2}, we describe the procedure for reconstructing $V(\phi)$ from $n_s(N)$.  
In Sec. \ref{sec3}, the case of $n_s-1=-2/N$ is studied. 
We shall find that known examples  exhaust the possibility:   for 
$n_s-1=-2/N$, $V(\phi)$ is 
either $\tanh^2(\gamma\phi/2)$ ("T-model") \cite{kl,kl1,kl2}
or  $\phi^2$ (chaotic inflation) to the leading order in the slow-roll approximation.  
We also examine the case of $n_s-1=-p/N$ in Sec.\ref{sec4}. 
We discuss the implications for the reheating temperature for the case of $n_s-1=-2/N$ in Sec.\ref{sec5}. 
Sec.\ref{secsum} is devoted to summary.

Related studies are given in  preceding works \cite{mukhanov,roest}. In \cite{mukhanov} the slow-roll parameter 
$\epsilon$ is given as a function of $N$ to construct $V(\phi)$. In \cite{roest} the slow-roll parameters $\epsilon$ and $\eta$ are given 
as functions of $N$ to construct $V(N)$ and compute $r$. 
Related results are found in \cite{garcia-roest,paolo}. In particular, $n_s-1=-p/N$ case is 
studied in \cite{paolo} by solving for the slow-roll parameter $\epsilon$. 
We study  the same problem by solving for the potential directly. 
Our approach is similar in spirit to \cite{mukhanov}; the only difference is that our starting point is one of the observables $n_s$ rather than 
the slow-roll parameter $\epsilon$. Moreover, we clarify the meaning of the integration 
constants and find a relation between $n_s(N)$ and $V(\phi)$.  

We use the units of $8\pi G=1$. 

\section{$V(\phi) $ from $n_s(N)$}
\label{sec2}

We explain the method to reconstruct $V(\phi)$ for a given $n_s(N)$. 
We study in the framework of a single scalar field with the canonical kinetic term 
coupled to the Einstein gravity. 
To do so, we first introduce the e-folding number $N$ and then the scalar spectral 
index $n_s$. 
The e-folding number $N$ measures the amount of inflationary  expansion  
from a particular time $t$ until the end of inflation $t_{\rm end}$
\beqa
N=\ln (a(t_{\rm end})/a(t))=\int^{t_{\rm end}}_{t}Hdt=\int^{\phi_{\rm end}}_{\phi}H\frac{d\phi}{\dot\phi}
=\int^{\phi_{\rm end}}_{\phi}\frac{H}{-V'/3H} d\phi
=\int^{\phi}_{\phi_{\rm end}}\frac{V}{V'}d\phi \,,
\label{N-phi}
\eeqa
where $\phi_{\rm end}=\phi(t_{\rm end})$ and the slow-roll equation of motion, 
$3H\dot\phi=-V'$,   
is used in the fourth equality. We assume that $N$ is large, say 
$N\sim O(10)\sim O(10^2)$, under  the slow-roll approximation. 
For the standard reheating process, $N\simeq 50\sim 60$ corresponds to 
the comoving scale $k$ probed by 
CMB experiments first crossed the Hubble radius during inflation ($k=aH$). 
In terms of the slow-roll parameters
\beqa
\epsilon\equiv \frac12\left(\frac{V'}{V}\right)^2,  \quad \eta\equiv\frac{V''}{V},  
\eeqa
$n_s$ is written as (to the first order in the slow-roll approximation),
\beqa
n_s-1=-6\epsilon +2\eta .
\label{ns:slow-roll}
\eeqa

The program to reconstruct $V(\phi)$ from $n_s(N)$ is to (i) first construct 
$V(N)$ from Eq. (\ref{ns:slow-roll}) 
and then to (ii) rewrite $N$ as a function of $\phi$ from Eq. (\ref{N-phi}) assuming large $N$. 
So, we first need to rewrite the slow-roll parameters as  functions of $N$. {}From Eq. (\ref{N-phi}), $dN=(V/V')d\phi$. 
Hence, we have
\beqa
V'=\frac{dV}{d\phi}=\frac{V}{V'}\frac{dV}{dN}\equiv \frac{V}{V'}V_{,N} .
\eeqa
Therefore, we obtain
\beqa
V'^2=VV_{,N} \,,
\label{V'}
\eeqa
{}from which $V_{,N}>0$ is required: $V$ is larger in the past in the slow-roll approximation.  
We note that 
this inequality also follows from $\dot H<0$ which holds as long as 
the weak energy condition is satisfied: 
\beqa
0>\dot H=-\frac12\dot\phi^2=-\frac16\frac{V'^2}{V}=-\frac{V_{,N}}{6},
\eeqa  
where the slow-roll equation of motion is used in the second equality. 

Assuming $V'>0$, from Eq. (\ref{V'}), we have
\beqa
V'=\left(VV_{,N}\right)^{1/2}
\eeqa
Similarly, we obtain
\beqa
V''=\frac{V}{V'}\frac{dV'}{dN}=\frac{V}{V'}\frac{d}{dN} \left(VV_{,N}\right)^{1/2}=
\frac{V_{,N}^2+VV_{,NN}}{2V_{,N}} .
\eeqa
Thus, Eq. (\ref{ns:slow-roll}) becomes
\beqa
n_s-1=-2\frac{V_{,N}}{V}+\frac{V_{,NN}}{V_{,N}}=\left(\ln\frac{V_{,N}}{V^2}\right)_{,N}  \, ,
\label{ns:N}
\eeqa
and $dN=(V/V')d\phi$ becomes
\beqa
\int \sqrt{\frac{V_{,N}}{V}}dN=\int d\phi \,.
\label{N-phi:2}
\eeqa
Eq. (\ref{ns:N}) and Eq. (\ref{N-phi:2}) are the basic equations for 
reconstructing $V(\phi)$ from $n_s(N)$. We also give the formulae for $r$ and 
the running of the spectral index:
\beqa
r&=&16\epsilon=8\frac{V_{,N}}{V},
\label{r:N}\\
\frac{dn_s}{d\ln k}&=&-\left(\ln\frac{V_{,N}}{V^2}\right)_{,NN}, 
\label{alpha:N}
\eeqa
where we have used $d\ln k=d\ln aH=-dN$ under the slow-roll approximation.

\section{$n_s-1=-2/N$}
\label{sec3}

As a warm-up, we consider the famous relation 
\beqa
n_s-1=-\frac{2}{N},
\label{ns:2/N}
\eeqa
which is in good agreement with the measurement of $n_s$ by Planck \cite{Planck:2015} 
for $N\simeq 60$.  We assume that Eq. (\ref{ns:2/N}) holds for  $N\gg 1$. 
Quadratic model, Starobinsky model, Higgs inflation and $\alpha$-attractors 
are known examples  which predict such a relation.   In this case, Eq. (\ref{ns:N}) becomes
\beqa
-\frac{2}{N}=\ln\left(\frac{V_{,N}}{V^2}\right)_{,N} ,
\eeqa
which is integrated to give
\beqa
\frac{V_{,N}}{V^2}=\frac{\alpha}{N^2},
\eeqa
where $\alpha$ is the integration constant and should be positive from $V_{,N}>0$. 
This equation is again integrated to obtain $V$
\beqa
V(N)=\frac{1}{\alpha/N+\beta},
\label{V:2}
\eeqa
where $\beta(\neq 0)$ is the second integration constant. 
The case of $\beta=0$ is to be considered separately. 
By taking the inverse of $V$, the meaning of the two integration constants is clear: 
$\beta$ is the value of $1/V$ at $N\rightarrow \infty$ and $\alpha$ is related to 
the value of $(1/V)_{,N}=-\alpha/N^2$.  

Given $V(N)$, we now proceed to the second step: rewrite 
$N$ as a function of $\phi$.  From Eq. (\ref{N-phi:2}), we have
\beqa
\int\sqrt{\frac{\alpha}{N(\alpha+\beta N)}}dN=\int d\phi,
\eeqa
which can be integrated depending on the sign of $\beta$. We first consider 
the case of $\beta>0$. For $\beta>0$, 
\beqa
N=\frac{\alpha}{\beta}\sinh^2\left(\sqrt{\beta/\alpha}(\phi-C)/2\right),
\eeqa
where $C$ is the integration constant corresponding to the shift  of $\phi$. 
Putting this into Eq. (\ref{V:2}), we finally obtain
\beqa
V(\phi)=\frac{1}{\beta}\tanh^2\left(\gamma(\phi-C)/2\right),
\eeqa
where we have introduced $\gamma=\sqrt{|\beta|/\alpha}$ which 
is the ratio of $1/V(N\rightarrow \infty)$ to $-N^2(1/V)_{,N}$. 
The potential is in fact the same as that of "T-model" \cite{kl}, as 
one might have expected, although $V$ is only accurate for large $\gamma\phi$ since 
we have used the slow-roll approximation and hence $N$ is large. So, $V$ is approximated as 
\beqa
V(\phi)\simeq \frac{1}{\beta}\left( 1-4 e^{-\gamma (\phi-C )}\right) .
\eeqa
The parameter $"\alpha"$ in the $\alpha$-attractor model \cite{kl} corresponds to $2/(3\gamma^2)$. The Starobinsky model corresponds to $\gamma=\sqrt{2/3}$. 

On the other hand, for $\beta<0$, 
\beqa
N=\gamma^{-2}\sin^2\left(\gamma(\phi-C)/2\right) .
\eeqa
$V(N)$ has a pole at $-\alpha/\beta=\gamma^{-2}$ 
and $N$ is restricted in the range $N<\gamma^{-2}$. Large $N$ is possible for small $\gamma$. 
Putting this into Eq. (\ref{V:2}), we obtain
\beqa
V(\phi)=-\frac{1}{\beta}\tan^2\left(\gamma(\phi-C)/2\right) .
\eeqa
However, from the slow-roll condition, 
\beqa
\epsilon=\frac{V_{,N}}{2V}=\frac{1}{2(N-\gamma^2N^2)}\ll 1,
\eeqa
$N\ll \gamma^{-2}$ is required. Therefore, the potential reduces to
\beqa
V(\phi)\simeq \frac{1}{4\alpha}(\phi-C)^2,
\eeqa
which is nothing but the quadratic potential. 

Finally, for $\beta=0$ (or $\gamma=0$) which corresponds to 
$V(N\rightarrow \infty)\rightarrow \infty$,  $V=N/\alpha$ and $N=(\phi-C)^2/4$, and so
\beqa
V(\phi)=\frac{1}{4\alpha}(\phi-C)^2,
\eeqa
which  precisely corresponds to the quadratic chaotic inflation model.

We also give  predictions for the tensor-to-scalar ratio $r$ and the running 
of the spectral index from Eq. (\ref{r:N}) and Eq. (\ref{alpha:N}): 
\beqa
r&=&
\left\{
\begin{array}{ll}
\frac{8}{N+\gamma^2 N^2}  & \mbox{($\beta> 0$)} \\
\frac{8}{N} &  \mbox{($\beta \leq 0$)}
\end{array}
\right.   
\label{r:2}
\\ 
\frac{dn_s}{d\ln k}&=&-\frac{2}{N^2}=-\frac12(1-n_s)^2\simeq -5.6\times 10^{-4}\left(\frac{60}{N}\right)^2 \,.
\eeqa
$r$ varies from $8/N$ (quadratic chaotic model) 
to $8/\gamma^2N^2$ (T-model) \cite{kl,kl1,kl2}. The measurement of $r$ could help to discriminate the model and 
to narrow down the shape of 
the potential. On the other hand, $dn_s/d\ln k$ does not depend on $\gamma$ since 
$n_s$ does not depend on it. $dn_s/d\ln k$ is definitely negative. 
Hence, the measurement of $dn_s/d\ln k$, which might be possible by  future observations of the 21 cm line \cite{oyama}, can be used as 
a consistency check of the assumed relation $n_s-1=-2/N$ \cite{paolo}.

\section{$n_s-1=-p/N$}
\label{sec4}
Next, we consider the more general relation 
\beqa
n_s-1=-\frac{p}{N},
\label{ns:p/N}
\eeqa
where $p>0$ and $p\neq 2$ are assumed. 
 First, from Eq. (\ref{ns:N}),  $V(N)$ is written as
\beqa
V(N)=\frac{1}{\frac{\alpha}{(p-1)}N^{1-p}+\beta}, 
\label{V:p}
\eeqa
where $\alpha(>0)$ and $\beta$ are the integration constants and we assume $p\neq 1$. 
We will consider the case of $p=1$ separately later. 
{}From Eq. (\ref{N-phi:2}), we have
\beqa
\int\sqrt{\frac{(p-1)\alpha}{(p-1)\beta N^p+\alpha N}}dN=\int d\phi,
\label{N-phi:3}
\eeqa
and the integration can be performed using the Gauss hypergeometric function, but 
the result is not illuminating. 
However, without using the hypergeometric function, 
we can see the asymptotic form of $V(\phi)$ for all cases of $\beta$ for large $N$. 

\subsection{$\beta>0$}
First, we consider several cases of $p$ for $\beta>0$. 
For $p>1$,  $N^p$ dominates over $N$ in Eq. (\ref{N-phi:3}) for large $N$ and the result is
\beqa
\phi-C\simeq -\frac{2}{(p-2)\gamma}N^{-(p-2)/2},
\label{N-phi:4}
\eeqa
where $\gamma=\sqrt{|\beta|/\alpha}$ and $p\neq 2$ is assumed.\footnote{For $p=2$, the integral gives $\ln N$ as given  in the previous section.  } 
$V(\phi)$ becomes for large $N$
\beqa
V(\phi)=\frac{1}{\beta}\left(1+\frac{\gamma^{-2}}{p-1}N^{1-p} \right)^{-1}\simeq 
\frac{1}{\beta}\left( 1-\frac{\gamma^{-2}}{p-1}\left(\left(1-\frac{p}{2}\right)\gamma(\phi-C)\right)^{2(p-1)/(p-2)}
\right). 
\label{V:p>1}
\eeqa
Although the functional form of $V(\phi)$ is the same, the behavior of $\phi$ for large $N$ is 
different  depending on whether $1<p<2$ or $p>2$ : 
For $1<p<2$, from Eq. (\ref{N-phi:4}), we find that $\phi$ increases as $N$ 
increases without bound, and $V(\phi)$ is of "Starobinsky" type (in the sense that the potential asymptotes to a constant from below 
for large $\phi$). 
On the other hand, for $p>2$, $\phi$ asymptotes to $C$ as $N\rightarrow \infty$, and 
$V(\phi)$ is of symmetry-breaking/hilltop type. 

Next, we consider the case $p<1$.  
In this case from Eq. (\ref{V:p}), $V(N)$ has a pole at  $N_*\equiv  
(\gamma^2(1-p))^{1/(1-p)}$ and $N$ is restricted in the range $N\ll N_*$. 
Large $N$ is   possible for large $\gamma$. 
Then,  $V_{,N}/V$ can be approximated as
\beqa
\frac{V_{,N}}{V}=\frac{1}{\gamma^2N^p}\frac{1}{1+\frac{\gamma^{-2}}{p-1}N^{1-p}}
=\frac{1}{\gamma^2N^p}\frac{1}{1- (N/N_*)^{1-p}}
\simeq \frac{1}{\gamma^2N^p} ,
\eeqa
assuming $N\ll N_*$. 
Therefore,  after all, the functional forms of $N(\phi)$ and $V(\phi)$ are  
the same  as Eq. (\ref{N-phi:4}) and Eq. (\ref{V:p>1}). Since the exponent of $\phi$ is $0<2(p-1)/(p-2)<1$ for $0<p<1$,  $V(\phi)$  may be called the "square-root" type.

\subsection{$\beta=0$}
For the case of $\beta=0$ and $p\neq 1$,  $V(N)=(p-1)N^{p-1}/\alpha$ from Eq. (\ref{V:p}) 
and hence $p>1$ is required. Then,  from Eq. (\ref{N-phi:3}), $N=(\phi-C)^2/4(p-1)$ and  
we obtain
\beqa
V(\phi)=\frac{p-1}{\alpha (4(p-1))^{p-1}}\left(\phi-C\right)^{2(p-1)},
\eeqa 
which is nothing but the  chaotic inflation model with a power-law potential.

\subsection{$\beta<0$}
For the case of $\beta<0$  and $p\neq 1$, a positive $V$ is possible only for $p>1$. In this case, 
from Eq. (\ref{V:p}), $V(N)$ has a pole at  
$N_*=  1/(\gamma^2(p-1))^{1/(p-1)}$ with $\gamma=\sqrt{|\beta|/\alpha}$ and 
$N$ is restricted in the range $N\ll N_*$. 
Large $N$ is   possible for small $\gamma$. 
Then,  $V_{,N}/V$ can be approximated as
\beqa
\frac{V_{,N}}{V}=\frac{p-1}{N}\frac{1}{1-{\gamma^{2}}(p-1)N^{p-1}}
\simeq \frac{p-1}{N} ,
\eeqa
assuming $N\ll N_*$. 
Hence, Eq. (\ref{N-phi:2}) is integrated to give
\beqa
N\simeq \frac{1}{4(p-1)}(\phi-C)^2 ,
\eeqa
and $V(\phi)$ can be written as
\beqa
V(\phi)\simeq -\frac{p-1}{\beta}N^{p-1}
\simeq \frac{p-1}{\alpha (4(p-1))^{p-1}}\left(\phi-C\right)^{2(p-1)}\,,
\eeqa
which is again the power-law potential.

\subsection{$p=1$}
Finally, we consider the case of $p=1$. In this case, 
from Eq. (\ref{ns:N})  $V(N)$ is written as
\beqa
V(N)=\frac{1}{\beta-\alpha\ln N},
\eeqa
where $\alpha$ and $\beta$ are the integration constants and are both positive.\footnote{$\beta$ here is no longer $1/V(N\rightarrow \infty)$ but $1/V(N=1)$. } So, $V(N)$ has a 
pole at$\ln N=\beta/\alpha\equiv \gamma^2$ and we can only consider the range 
$\ln N\ll \gamma^2$. Large $N$ is possible for large $\gamma$. 
Then, $V_{,N}/V$ can be approximated as
\beqa
\frac{V_{,N}}{V}=\frac{1}{\gamma^2 N-N\ln N}\simeq \frac{1}{\gamma^2 N},
\eeqa
assuming $\gamma$ is large. 
Hence, Eq. (\ref{N-phi:2}) is integrated to give
\beqa
N\simeq \frac{\gamma^2}{4}(\phi-C)^2 ,
\eeqa
and $V(\phi)$ can be written as
\beqa
V(\phi)=\frac{1}{\beta}\frac{1}{1-\gamma^{-2}\ln N}
\simeq \frac{1}{\beta}\left(1+2\gamma^{-2}\ln (\gamma (\phi-C)/2)\right) ,
\eeqa
and $V(\phi)$ is of logarithmic type.

\begin{figure}
\includegraphics[height=4in]{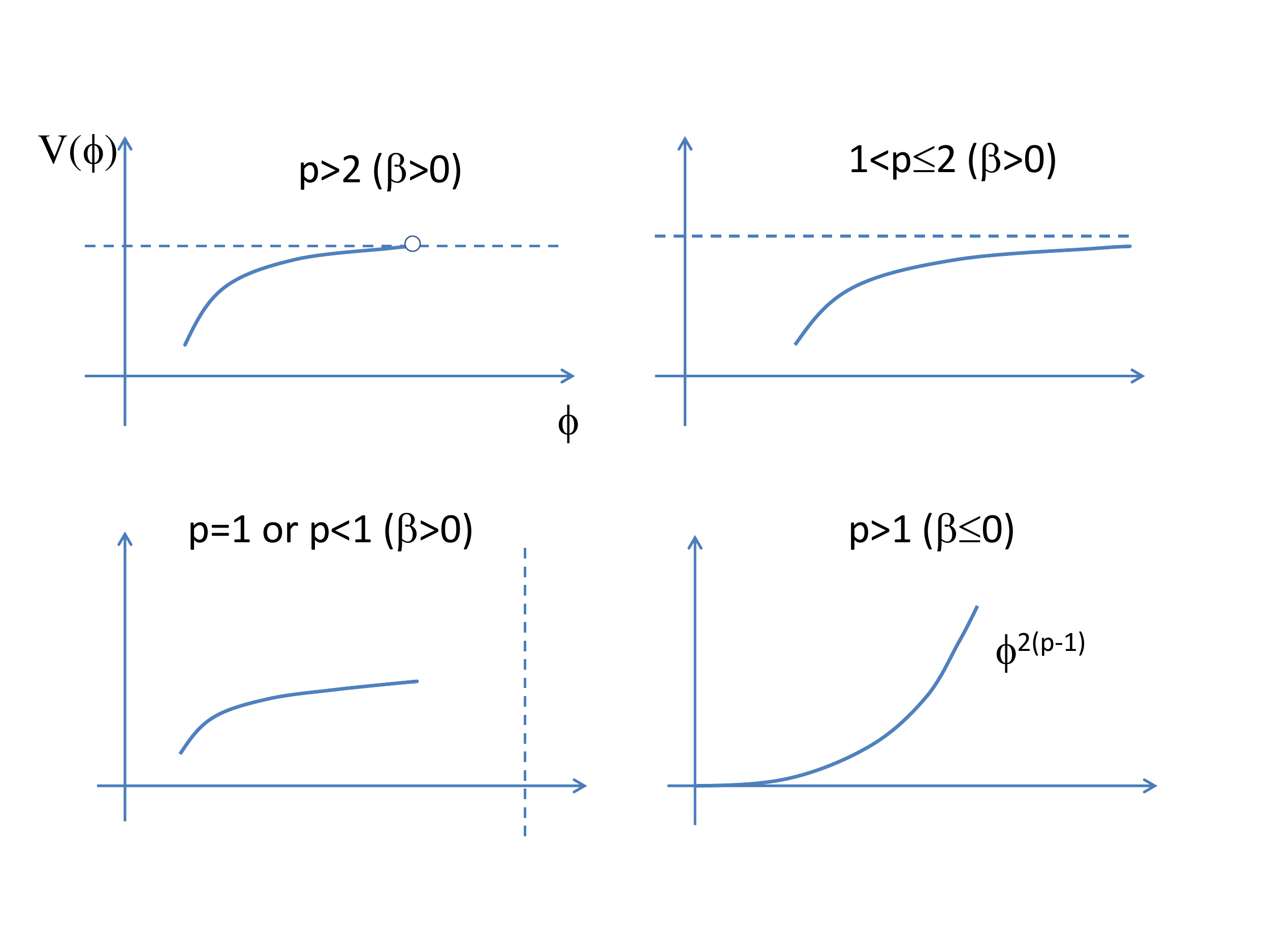}
\caption{\label{fig1} The schematic shape of $V(\phi)$ for several cases of $p$.  
$\beta=1/V(N\rightarrow \infty)$.  
 }
\end{figure}

The schematic shape of the potential for each case of $p$ is shown in Fig. \ref{fig1}. 
We find that if $V(\phi)$ is bounded from above, only $p>1$ is allowed. 

We also give the predictions for the tensor-to-scalar ratio $r$ and the running 
of the spectral index from Eq. (\ref{r:N}) and Eq. (\ref{alpha:N}): 
\beqa
r&=&
\left\{
\begin{array}{ll}
\frac{8(p-1)}{N+\gamma^2(p-1) N^p} & \mbox{($\beta> 0$)} \\
 \frac{8(p-1)}{N}  & \mbox{($\beta\leq 0$)} \\
 \frac{8}{\gamma^2 N}  &  \mbox{($p=1$)}
\end{array}
\right.   
\\
\frac{dn_s}{d\ln k}&=&-\frac{p}{N^2} \,.
\eeqa
$r$ varies from $8(p-1)/N$ (chaotic inflation model) 
to $8/\gamma^2N^p$ (modified T-model).

\section{Reheating Temperature} 
\label{sec5}

Once  $V(N)$ is given, it is possible to connect $N$ and the reheating temperature 
$T_{\rm RH}$ assuming that $V(N)$ is valid for smaller $N$ \cite{mark1,kuro,mark2,cook}. \footnote{We would like to thank 
Kazunori Kohri for suggesting this point. }
For simplicity we consider the case of $n_s-1=-2/N$. 

For the mode with wavenumber $k$, the comoving Hubble scale when this mode exited the horizon ($k=aH$) is related to that of the present time, $a_0H_0$ by
\beqa
\frac{k}{a_0H_0}=\frac{H}{H_0}\frac{a}{a_{\rm end}}
\frac{a_{\rm end}}{a_{\rm RH}}\frac{a_{\rm RH}}{a_0},
\label{k/N}
\eeqa
where $a_{\rm end}=a(t_{\rm end})$ and $a_{\rm RH}$ is the scale factor at the end of 
reheating. Here, by definition, $a/a_{\rm end}=e^{-N}$. 
Moreover, assuming that during the reheating phase 
the effective equation of state of the universe is matter-like due to coherent oscillation 
of the inflaton,  the energy density at the end of inflation $\rho_{\rm end}$ is related to 
the energy density at the end of reheating $\rho_{\rm RH}$  by
$\rho_{\rm RH}/\rho_{\rm end}=(a_{\rm end}/a_{\rm RH})^3$. $\rho_{\rm RH}$ is 
$\rho_{\rm RH}=(\pi^2g_{\rm RH}/30)T_{\rm RH}^4$, where $g_{\rm RH}$ is the effective number 
of relativistic degrees of freedom at the end of reheating. 
Further, assuming the conservation of entropy, $g_{S,\rm RH}a_{\rm RH}^3T_{\rm RH}^3=(43/11)a_0^3T_0^3$, where $g_{S,\rm RH}$ is the effective number 
of relativistic degrees of freedom for entropy at the end of reheating. 
Finally, the Hubble parameter during inflation is related to the scalar amplitude $A_s$ by 
$H^2=V/3=(\pi^2/2)r A_s$, where $A_s\simeq 2.14\times 10^{-9}$ from Planck 
\cite{Planck:2015}. Plugging these relations into Eq. (\ref{k/N}), we obtain \cite{kuro}
\beqa
N=56.9-\ln\frac{k}{a_0H_0}-\ln\frac{h}{0.67}-\frac13\ln\frac{\rho_{\rm end}}{\rho(N)}
+\frac13\ln\frac{T_{\rm RH}}{10^9{\rm GeV}}+\frac16 \ln r(N) ,
\eeqa
where $h$ denotes the dimensionless Hubble parameter and we have set $g_{\rm RH}=g_{S,\rm RH}$. From the requirement that the energy density at the end of reheating should be smaller than the energy density at the end of inflation, we have an upper bound on the reheating temperature $T_{\rm crit}$
\beqa
\frac{\pi^2g_{\rm RH}}{30}T^4< \frac{\pi^2g_{\rm RH}}{30}T_{\rm crit}^4=\rho_{\rm end}=\frac{\rho_{\rm end}}{V}\frac{3\pi^2}{2}r A_s .
\label{Tcrit}
\eeqa

For the potential Eq. (\ref{V:2}), the ratio is estimated as
\beqa
\frac{\rho_{\rm end}}{\rho(N)}\simeq \frac43\frac{V_{\rm end}}{V(N)}=
\frac43\left(\frac{\sqrt{1+2\gamma^2}-1}{\sqrt{1+2\gamma^2}+1    } \right)
\left(\frac{1+\gamma^2N}{\gamma^2N}\right),   
\eeqa
where the factor $4/3$ comes from the contribution of the inflaton 
kinetic energy to $\rho_{\rm end}$ and we have defined the end of inflation by $\epsilon=V_{,N}/2V=1$. 
Therefore, $T_{\rm RH}$ can be written as
\beqa
\frac{T_{\rm RH}}{10^9 {\rm GeV}}=  e^{3(N-56.9)}\left(\frac{k}{a_0H_0}\right)^3\frac43
\left(\frac{\sqrt{1+2\gamma^2}-1}{\sqrt{1+2\gamma^2}+1    } \right)
\left(\frac{1+\gamma^2N}{\gamma^2N}\right)
\left(\frac{N+\gamma^2N^2}{8}\right)^{1/2} ,
\eeqa
where we have used Eq. (\ref{r:2}) and assumed $h=0.67$. 
The upper bound of the reheating temperature 
Eq. (\ref{Tcrit}) becomes
\beqa
T_{\rm crit}=1.43\times 10^{16}{\rm GeV}\left(\frac{\sqrt{1+2\gamma^2}-1}{\sqrt{1+2\gamma^2}+1    } \right)^{1/4}
\left(\frac{1+\gamma^2N}{\gamma^2N}\right)^{1/4}
\left(\frac{N+\gamma^2N^2}{8}\right)^{1/4},
\label{Tcrit2}
\eeqa
where we have assumed $ g_{\rm RH}=106.75$. 
\begin{figure}
\includegraphics[height=4in]{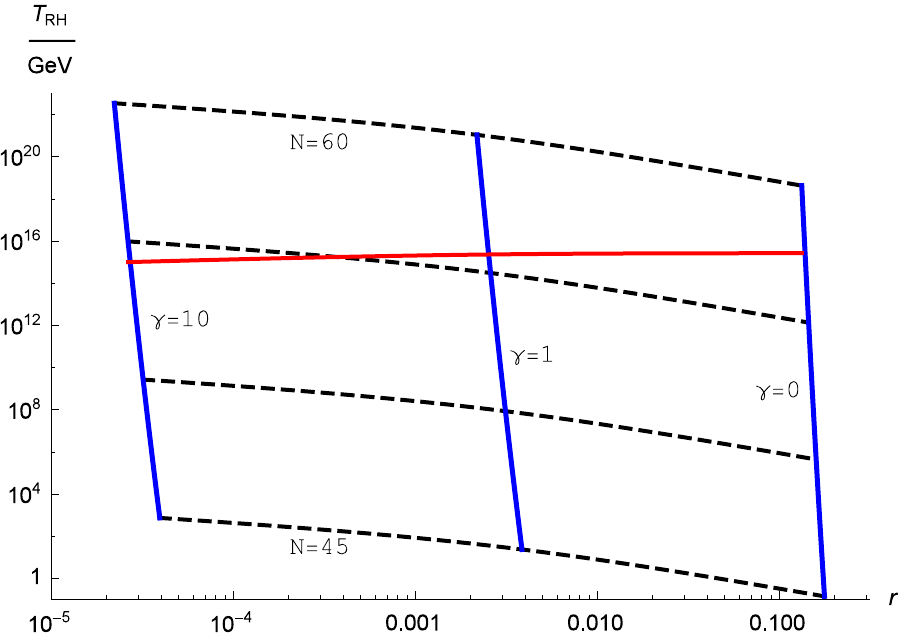}
\caption{\label{fig2} The reheating temperature versus $r$ for the model $n_s-1=-2/N$ 
for several $\gamma$ and $N$. The pivot scale is $k=0.05{\rm Mpc}^{-1}$.  
 Blue curves for $\gamma=10,1,0$ from left to right; dotted curves for $N=60,55,50,45$ from top to bottom. The red curve is the upper bound on the reheating temperature $T_{\rm crit}$.
 }
\end{figure}
\begin{figure}
\includegraphics[height=4in]{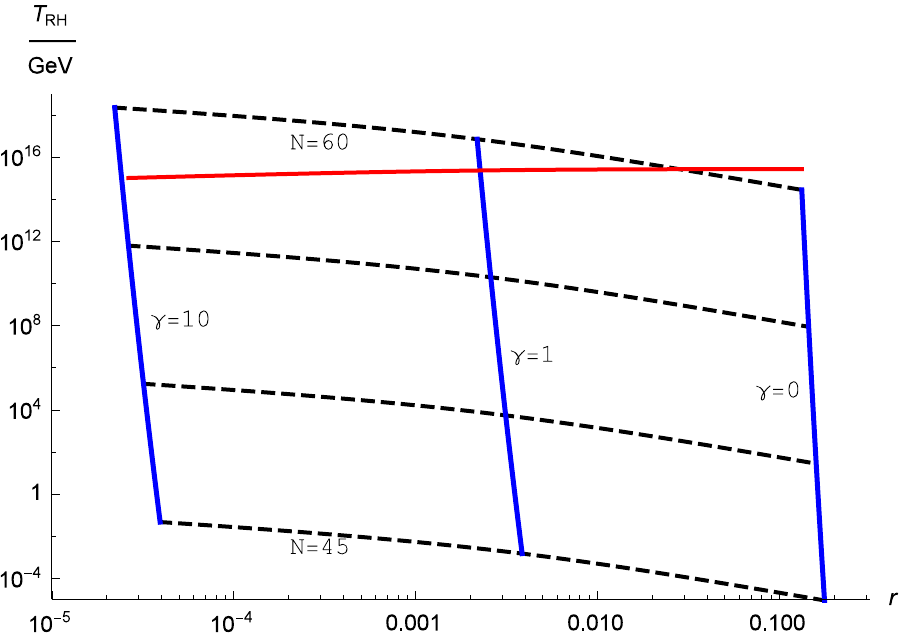}
\caption{\label{fig3} Same as Fig. \ref{fig2} but for the pivot scale $k=0.002{\rm Mpc}^{-1}$.
 }
\end{figure}

In Fig. \ref{fig2}, we show the reheating temperature versus $r$ for several $\gamma$ and $N$ 
for the pivot scale $k=0.05{\rm Mpc}^{-1}$ at which Planck determines $n_s$ \cite{Planck:2015}.
The blue curves are for $\gamma=10,1,0$ from left to right and dotted curves are for $N=60,55,50,45$ from top to bottom. For $n_s-1=-2/N$, the Planck result $n_s=0.9688\pm 0.0061$ \cite{Planck:2015} implies $53.8<N<79.7$. 
We find that for fixed $N$ $T_{\rm RH}$ slightly increases as $\gamma$ increases. 
This is because for larger $\gamma$ the slow-roll parameter $\epsilon$ becomes smaller 
and the Hubble parameter decreases less (in time)  and hence the comoving 
Hubble horizon at the end of inflation $1/a_{\rm end}H_{\rm end}$ becomes smaller 
and this results in the  shorter duration of the reheating phase. 

The red curve is the upper bound on the reheating temperature $T_{\rm crit}$, Eq. (\ref{Tcrit2}). 
Since the dependence of  $T_{\rm crit}$ on $N$ is very weak, we 
show the curve for $N=60$. The upper bound on the reheating temperature 
puts the upper bound on $N$: $N\siml 57$ for $\gamma=0$,  $N\siml 56$ for $\gamma=1$ 
and $N\siml 54$ for $\gamma=10$.   The result for $\gamma=0$ is consistent with \cite{mark2}. 
Combined with the measurement of $n_s$, $\gamma$ is constrained as $\gamma\siml 10$. 
Further combined with the constraint on $r<0.12$ by BICEP2/Planck \cite{bicep}, 
we obtain $0.1\siml \gamma\siml 10$ and $10^{11}{\rm GeV}\siml T_{RH}\siml 2\times 
 10^{15}{\rm GeV}$.  We note that the reheating temperature depends strongly on 
the equation of state during the reheating stage \cite{mark1}. 
The upper bounds on $N$ are significantly relaxed for the pivot scale $k= 0.002{\rm Mpc}^{-1}$ 
because $T_{\rm RH}$ depends on $k^3$ (see Fig. \ref{fig3}).  

%
\section{Summary} 
\label{secsum}

Motivated by the relation  $n_s-1\simeq -2/N$ indicated by recent cosmological observations, we derive the formulae to  
derive the inflaton potential  $V(\phi)$ from $n_s(N)$. Applied to $n_s-1=-2/N$, to the first order in the slow-roll approximation,  
we find that the potential is classified into two categories depending on the value of  $1/V(N\rightarrow \infty)$:  
T-model type ($\tanh^2(\gamma\phi/2)$) for $1/V(N\rightarrow \infty)> 0$ 
or  quadratic type ($\phi^2$) for $1/V(N\rightarrow \infty)\leq 0$. 
$\gamma$ is the ratio of $1/V(N\rightarrow \infty)$ to $-N^2(1/V)_{,N}$. 
We have calculated the reheating temperature versus the tensor-scalar ratio diagram. 
We have found that for fixed $N$ the reheating temperature slightly increases as $\gamma$ increases. 
For the pivot scale $k=0.05{\rm Mpc}^{-1}$, the upper bound on 
the reheating temperature puts the upper bound on the e-folding number, $N<60$. 

We extend the classification of the potential $V(\phi)$ for $n_s-1=-p/N$. The shape of the potential is classified into 
four categories: symmetry-breaking type ($p>2$ and $1/V(N\rightarrow \infty)> 0$),    
Starobinsky type ($1<p\leq 2$ and $1/V(N\rightarrow \infty)> 0$),  
square-root/logarithmic type ( $0<p<1$ and $1/V(N\rightarrow \infty)> 0$ or $p=1$),  
and  power-law type $\phi^{2(p-1)}$ for $1/V(N\rightarrow \infty)\leq 0$.  
We find that  only $p>1$ is allowed  for the potential bounded from above. 

We find that although $r$ depends on the ratio of the integration constant $\gamma$, 
the running of the spectral index does not (by construction). 
Therefore, the measurement of $r$ can be used to discriminate the model and 
to narrow down the shape, while  the measurement of the running is used as a consistency check of the assumed form of $n_s(N)$.

\section*{ACKNOWLEDGEMENTS}

We would like to thank  
Kazunori Kohri and Masahide Yamaguchi 
for useful comments. 
This work is supported by the Grant-in-Aid for Scientific Research
from JSPS (Nos.\,24540287) and in part
by Nihon University. 


\end{document}